\documentclass[]{spie}  

\usepackage{amsmath,amsfonts,amssymb}
\usepackage{graphicx}
\usepackage[colorlinks=true, allcolors=blue]{hyperref}
\usepackage{siunitx}
\usepackage{orcidlink}
\usepackage{caption}
\usepackage{subcaption}
\usepackage{todonotes}

\title{Performance and Applications of Optical Pin Beams in Turbulent Long-Range Free Space Optical Communications}

\author[a,d]{Francesco Nardo \orcidlink{0009-0006-5750-5837} }
\author[b]{Jan Tepper}
\author[c]{Ricardo Barrios \orcidlink{0000-0002-2545-5568} }
\author[d]{Jonas Krimmer \orcidlink{0009-0002-7005-0428} }
\author[d]{\\ Sebastian Randel}

\affil[a]{Central Research and Technology, Airbus DS GmbH, Taufkirchen, Germany}
\affil[b]{Space Systems, Airbus DS GmbH, Taufkirchen, Germany}
\affil[c]{System Engineering SATCOM Air comms, Airbus DS GmbH, Taufkirchen, Germany}
\affil[d]{Institute of Photonics and Quantum Electronics, Karlsruhe Institute of Technology, Karlsruhe, Germany}

\authorinfo{Further author information: (Send correspondence to F.Nardo)\\F.Nardo: \ \ E-mail: fn.francesconardo@gmail.com, \\  R.Barrios: \ E-mail: ricardo.barrios@airbus.com,  \\ S.Randel: \ E-mail: sebastian.randel@kit.edu}

\usepackage{eso-pic}
\usepackage{tcolorbox}
\AddToShipoutPicture*{%
  \AtPageUpperLeft{%
    \raisebox{-1.5cm}{ 
      \makebox[\paperwidth]{%
        \centering
        \begin{minipage}{1\textwidth} 
          \centering
          \footnotesize This paper has been accepted for poster presentation at SPIE Photonics West 2025. ©2025 SPIE.
          Please cite it as: \\
          F. Nardo, J. Tepper, R. Barrios, J. Krimmer, and S. Randel, 
          "Performance and applications of optical pin beams in turbulent long-range free-space optical communications,"
          \textit{Proc. SPIE 13355, Free-Space Laser Communications XXXVII, 133551H (19 March 2025)}; 
          \url{https://doi.org/10.1117/12.3040720}
        \end{minipage}
      }
    }
  }
}
\AddToShipoutPicture{ 
  \AtPageLowerLeft{%
    \raisebox{0.5cm}{ 
      \makebox[\paperwidth]{%
        \begin{tcolorbox}[colframe=black, colback=white, sharp corners, width=\textwidth, ]
          \footnotesize ©2025 Society of Photo‑Optical Instrumentation Engineers (SPIE). One print or electronic copy may be made for personal use only. Systematic reproduction and distribution, duplication of any material in this publication for a fee or for commercial purposes, and modification of the contents of the publication are prohibited.
        \end{tcolorbox}
      }
    }
  }
}
 
\begin{document}

\maketitle

\begin{abstract}
Optical pin beams (OPBs) are a promising candidate for realizing turbulence-resilient long-distance free-space optical communication links spanning hundreds of kilometers. 
In this work, we introduce a unified theoretical model to describe the propagation of OPBs and present comprehensive simulation results based on many realizations and link-budget analyses for constant turbulence strengths.
For reference, we compare the performance of the OPBs to weakly diverging and focusing Gaussian beams.
For a \qty{100}{\km} long air-to-air link, \qty{10}{\km} above sea level, our simulation results show that OPBs offer an improved link budget of up to \qty{8.6}{\dB} and enhanced beam wander statistics of up to \qty{3}{\dB} compared to the considered Gaussian beams.
Additionally, we identified a quadratic relationship between the transmitter aperture diameter and the maximum achievable distances, which is crucial in deciding the suitability of OPBs for a given application scenario.
\end{abstract}


\section{INTRODUCTION}
\label{sec:intro}  
In air-to-air communication scenarios, free-space optical (FSO) communication offers extremely high bandwidth, unlicensed spectrum allocation, reduced power consumption, reduced antenna dimensions and improved channel security \cite{Chan_book} compared to traditional radio-frequency channels \cite{Kaushal_survey_2017}. 
%
%
In this context, non-diffracting, structured light has been proposed as an alternative to Gaussian beams (GBs). 
In particular, optical pin beams (OPBs) were first introduced in 2019 by Zhang et al. \cite{zhang_robust_2019}, where the authors demonstrated that a GB, traversing an etched phase mask with a power-law profile, can result in a stable optical field, exhibiting “self-focusing” dynamics. 
In Hu et al. \cite{hu_experimental_2022}, the authors experimentally demonstrated the benefits (in terms of link-budget) of using an OPB instead of a collimated GB for transmitting a 
\qty{1}{\giga \bit \per \s} on-off-keying signal over a \qty{1}{m} FSO channel. 
In 2020, Li et al. \cite{li_direct_2020} expanded the work by Zhang et al. by generalizing the OPB theoretical model to a new class of beams where
the power-law has an exponent $\gamma$ which can physically take values within the range of \qty{0}{} to \qty{2}{}. When $\gamma$ is \qty{1}{}, a Bessel beam (BB) is formed after initial propagation, while $\gamma$ equal to \qty{1.5}{} yields the specific solution previously introduced in Zhang et al.\cite{zhang_robust_2019}.
Yu et al. \cite{yu_self-healing_2023} experimentally demonstrated and discussed the self healing abilities of this new, more general theoretical description of OPBs.
Finally, a further generalization follows in 2023 by Droulias et al.\cite{droulias_inverted_2023}, where the source field is the product between an arbitrary radially symmetric function $B(r)$ and the previously introduced power-law. 
Related works have been focusing mainly on \qty{532}{\nm} (visible light) in experimental setups regarding OPBs \cite{zhang_robust_2019,li_direct_2020,yu_self-healing_2023,droulias_inverted_2023}. As we will later demonstrate, given the same transmitting aperture, the use of a shorter wavelength is advantageous.
However, in most FSO applications, longer wavelengths (\qtyrange[range-phrase=\,--\,]{500}{2000}{\nm}) \cite{Kaushal_survey_2017} are preferred due to the transmission, absorption and scattering windows through atmosphere\cite{hemmati_book}.

In particular, in high bandwidth FSO, as in Hu et al.\cite{hu_experimental_2022}, $\lambda = \qty{1550}{\nm}$ has been chosen as the operational wavelength after considering the availability of highly optimized, telecoms equipment.

This paper concentrates on Li et al.'s general solution \cite{li_direct_2020}.
OPBs are generated from a Gaussian source, followed by numerical modeling of their propagation through the atmospheric channel using the angular-spectrum method.
Assuming constant clear-air (excluding atmospheric absorption and scattering effects) turbulence at a constant altitude of \qty{10}{\km} for a \qty{100}{\km} link path, we compare in terms of scintillation index and power received by a \qty{10}{\cm} aperture, the statistical results from many realizations of: 1) a collimated GB i.e. a slowly diverging GB with wavefront radius of curvature $R \rightarrow \infty$, 2) a focused GB with $R = \qty{100}{\km}$ and 3) various OPBs with different power-law exponents. 
While the characteristics of OPBs have been numerically studied in previous publications (and validated through experimental setups), to our knowledge, realistic simulations of distances that are relevant in aerospace FSO applications, i.e., hundreds of kilometers, have not been addressed yet.



\begin{figure*}[t]
    \centering
    \includegraphics[width=1\textwidth]{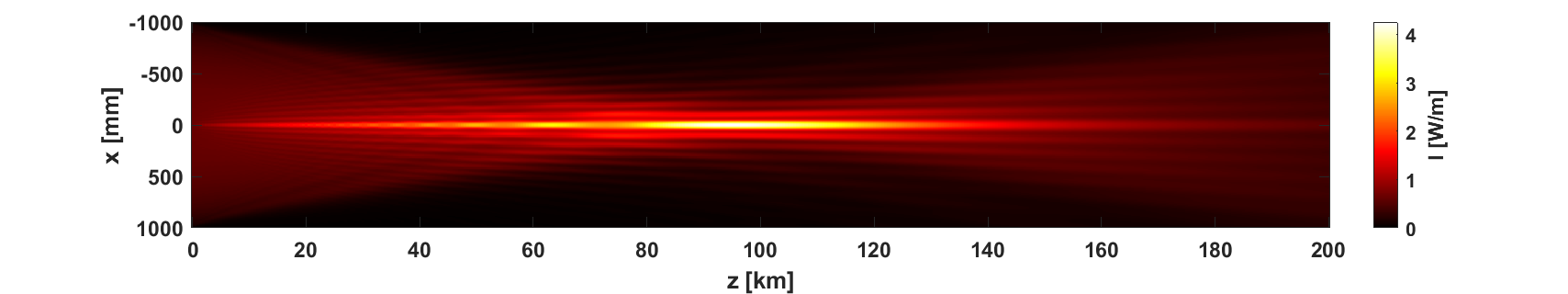}
    \caption{Intensity profile as function of distance of an OPB in vacuum with $\gamma=$ \qty{0.1}{}, $z_m =$ \qty{125}{\km}, $A_{TX}=$ \qty{2}{\m}, $P_{TX}=$ \qty{1}{\watt}. }
    \label{fig:intensity_profile}
\end{figure*}

\section{Theoretical Model}
We consider a beam propagating along the $z$-direction with wavenumber $k=2\pi/\lambda$ , and denote the radial coordinate in the transverse ($x, y$) plane as $\smash{r=\sqrt{x^2 + y^2}}$.
The theoretical definition of a beam belonging to the OPB class has been provided by Li et al.\cite{li_direct_2020}.
Reportedly, after a certain distance from the source plane, the amplitude assumes the shape of a zeroth-order Bessel function.
The full width at half-maximum (FWHM) of OPBs is associated with the central peak of this Bessel function and it is defined as\cite{droulias_inverted_2023}:
\begin{equation} 
     \mathrm{FWHM}(z) = 2.27 \cdot k^{-1} \cdot (C \gamma z^{(\gamma-1)})^{\frac{1}{2-\gamma}}
 \label{eq:fwhm_theo}
\end{equation}
The constant $C$ depends on the maximum propagation distance $z_m$, after which an OPB starts to rapidly diverge, and the diameter of the transmitting aperture $A_{TX}$:
\begin{equation}
 C = \frac{(A_{TX}/2)^{2-\gamma}}{\gamma z_m }
 \label{eq:C}
\end{equation}
The (planar) wavefront of a perfectly collimated GB, can be modulated to match the phase-profile of an OPB by using a Spatial Light Modulator (SLM)\cite{yu_self-healing_2023}. 
The complex field of such a beam, in the transverse source plane, can be expressed as:
\begin{equation}
    \psi(r, z = 0) = \psi_0 \cdot A(r) \cdot \mathrm{e}^{-\frac{r^2}{w_0^2}} \cdot \mathrm{e}^{\mathrm{-j}  \cdot k C r^\gamma}
    \label{eq:OPB_input}
\end{equation}
where $\psi_0$ is simply a constant amplitude factor, $w_0$ is the beam waist and $A(r)$ describes a circular transmit aperture with diameter $A_{TX}$.


\section{Numerical Model}
OPBs represent an asymptotic solution to the paraxial Helmholtz equation and the Fresnel propagation integral can be used to simulate their propagation in vacuum\cite{li_direct_2020}. The widely adopted approach involves expressing the Fresnel integral as a convolution and numerically evaluating it using Fourier Transforms, a method known as the angular-spectrum method \cite{schmidt_numerical_2010}.
To simulate clear-air turbulence effects for a \qty{100}{\km} horizontal link at an altitude of \qty{10}{\km},
we assumed a constant structure coefficient\cite{Kolmogorov_1941} $C_n^2 = \qty[exponent-product=\cdot]{1.66e-17}{\m^{-2/3}}$ according to the HV-5/7 model\cite{stotts_HAP_2023_improving}.
The turbulent phase screens are generated using the Fourier method with added sub-harmonics as introduced by Lane et al.\cite{lane_simulation_1992}.
We are using the modified von Kármán power spectral model\cite{andrews_book_2005}, where the inner and outer scales are $l_0 = $ \qty{1}{\cm} and $L_0 = \infty$ respectively and $N_{sh} = 5$ sub-harmonics have been added\cite{carbillet_N_subharmonics_2010,Jonas_2020}. 
The computational grid consists of $1024 \times 1024$ pixels providing a spatial resolution of \qty{4.9}{mm}. 
The quadratic domain dimensions are $5\times5$ \qty{}{\metre\squared}, a size required to ensure that all energy remains contained within the propagation grid until reaching the receiver, in particular for diverging ($\gamma < 1$) OPBs. The diameter of the transmitting aperture is $A_{TX} = \qty{2}{\m}$, a large value necessitated by OPB requirements, as will be shown later. 
In total, there are \qty{21}{} planes, spaced every \qty{5}{\km} until reaching the receiver, in addition to the input plane.
The equal spacing between propagation planes has been selected to operate within the near field condition, meeting grid and propagation geometry constraints, and finally to limit the Rytov number between two successive planes $\sigma_R = 0.08 < 0.1$.  When considering the full link path length, the Rytov number is $\sigma_R = 1.24$, i.e., corresponding to the strong fluctuation regime\cite{andrews_book_2005}.

Figure \ref{fig:intensity_profile} illustrates the intensity profile of an OPB realization in vacuum using the angular-spectrum method through \qty{400}{} planes (to increase the $z$-resolution of the profile), with the source plane defined by Eq. \eqref{eq:OPB_input}. 
The beam waist is $w_0 = \qty{1}{\m}$, $A_{TX} = \qty{2}{\m}$ and $\psi_0$ is chosen such that the total power (integration of intensity over transmitting aperture) is $P_{TX}=\qty{1}{\watt}$. 

\section{Minimum Transmitting Aperture}
\begin{figure}[ht!]
    \centering
    \includegraphics[width=0.95\textwidth]{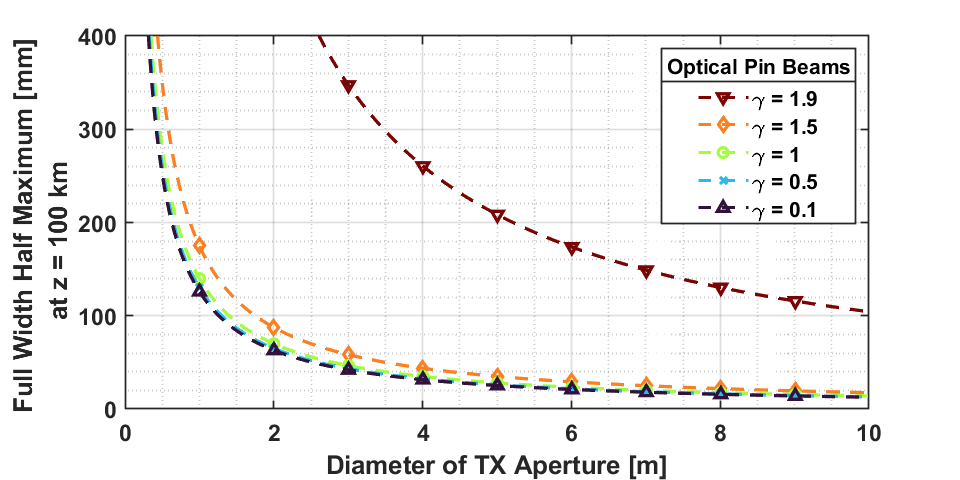}
    \caption{Full Width Half Maximum at the plane $z=$ \qty{100}{\km} for an OPB with $z_m =$ \qty{125}{\km} as function of Aperture Diameter $A_{TX}$. \\ }
    \label{fig:ring_optimization}
\end{figure}

Non-diffracting beams, while theoretically feasible, require an infinite aperture size due to their infinite energy content \cite{Cox_bessel_beam_1992}. In practical scenarios, a finite aperture size is utilized, so it is the scope of this article to determine the minimum diameter needed for the propagation of OPBs.  
Figure \ref{fig:ring_optimization} illustrates the relationship between the minimum achievable FWHM at $z=$ \qty{100}{\km} for an OPB, Eq. \eqref{eq:fwhm_theo}, with the numerical findings for fixed $z_m = $ \qty{125}{\km}, as function of the aperture diameter for different $\gamma$ values. 
Since $z_m$ represents the maximum propagation distance of an OPB, it must be greater than the intended target distance, ensuring that the receiver falls within the focusing region.


In analogy with focused GBs, where the focal length is proportional to the curvature of the wavefront and therefore large curvatures are required to focus at great distances, OPBs exhibit what is known as holographic periods\cite{Zhai_SLM_axicon_2020}, where the maximum propagation distance is proportional to the holographic period. 
If the aperture diameter is reduced beyond the necessary holographic period for a given distance, the formation or propagation of the OPB becomes impossible. Conversely, increasing the aperture includes more periods, and the properties of OPBs, such as FWHM, rapidly improve up to a certain limit (observed as the knee of the curve in Figure \ref{fig:ring_optimization}).
Interestingly, in our simulation, we observed that when using the same aperture (beam truncation) and wavefront profile, a larger beam waist of the Gaussian amplitude (similar to a top-hat beam) results in a smaller FWHM at the target distance.
This observation may be explained by the generation of the focusing region for OPBs, attributed to the destructive interference of transverse wavevector components and constructive interference for the longitudinal ones \cite{zhang_robust_2019}. Consequently, with an uniform field amplitude underlying the wavefront, the contribution from outer holographic periods is equally weighted. 

\begin{figure}[ht!]
    \centering
    \includegraphics[width=1\textwidth]{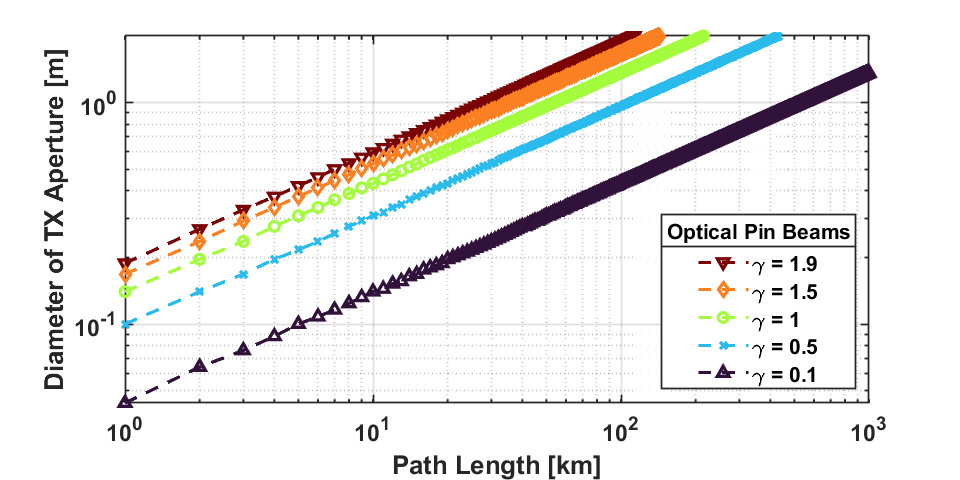}
    \caption{Engineering chart for $\lambda= \qty{1550}{\nano \m}$. In log-log scale the minimum aperture required has linear proportionality with the path length. When $\gamma = \qty{1.9}{}$ the \qty{1}{\km} link requires a minimum aperture for the transmitting optics of \qty{18.8}{\cm} while the \qty{100}{\km} link requires \qty{1.88}{\m}. \\ }
    \label{fig:protype_diagram}
\end{figure}

When considering different link lengths, the required aperture size to transmit at least three holographic periods (the curve's inflection point) is summarized in the plots of Figure \ref{fig:protype_diagram}.
Notably, using a smaller $\gamma$, or shorter $\lambda$, reduces the required diameter of the transmitting aperture. For BB ($\gamma = 1$) propagating on a \qty{100}{\km} path length, the minimum aperture diameter required is \qty{1.4}{\m}.
From fitting the so obtained data, it emerges that the proportional relationship between maximal achievable distance (desired path length) and aperture diameter is quadratic, i.e. $z_{max} \propto A_{TX}^2$. For example, when $\gamma = \qty{1.9}{}$ the \qty{1}{\km} link requires a minimum aperture for the transmitting optics of \qty{18.8}{\cm} while the \qty{100}{\km} link requires \qty{1.88}{\m}, i.e. an aperture that is \qty{10}{} times larger allows a path length that is \qty{100}{} times longer. 

\section{Link Budget Improvements}
\begin{figure}[ht!]
    \centering
    \includegraphics[width=0.9\textwidth]{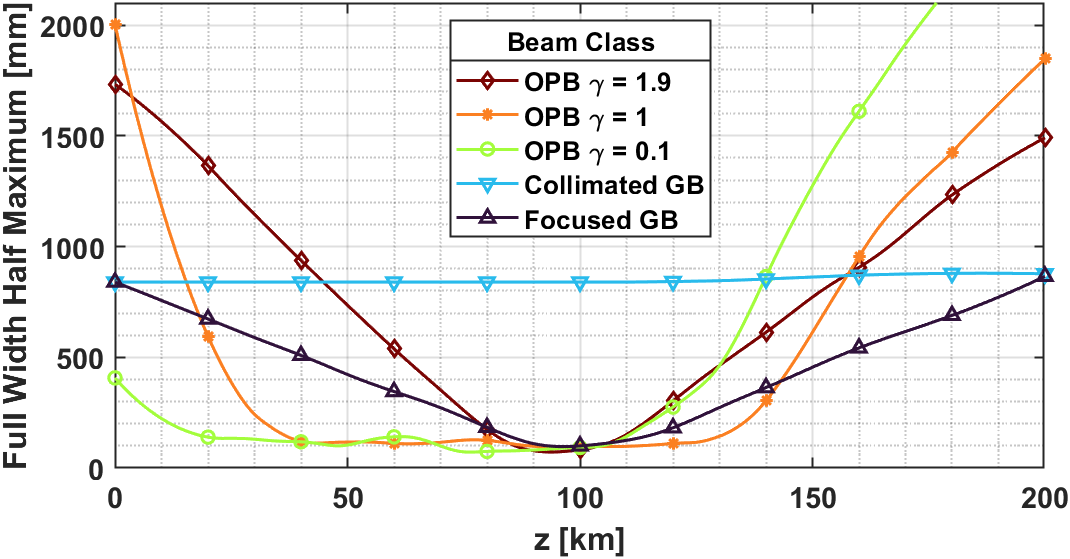}
    \caption{FWHMs of the intensity profile in vacuum. Comparison between: collimated GB, focused GB with $w_0 = A_{TX}/\sqrt{8} = $ \qty{0.7}{\m} and OPBs with $z_m = $ \qty{125}{\km}. All beams have the same $A_{TX} = $ \qty{2}{\m} and $P_{TX} =$ \qty{1}{\watt}. \\ \medskip}
    \label{fig:fwhm}
\end{figure}

The FWHMs of different beam types, in vacuum, are plotted against propagation distance in Figure \ref{fig:fwhm}.
Compared to the GBs, OPBs allow to maintain a smaller FWHM in the focus region for an extension of tens of kilometers.
For the same diameter of the receiving aperture $A_{RX}$, this can result in an improved link budget (received power $P_{RX}$) in particular for a moving target.

In Table \ref{tab:statistics} we have summarized over many realizations, the average beam wander $r_c$, average power received $P_{RX}$ and scintillation index $\sigma_I^2$ for various beams, all featuring identical settings: $A_{TX}$, $P_{TX}$, $A_{RX}$, while subject to a constant turbulence strength of $C_n^2 = \qty[exponent-product=\cdot]{1.66e-17}{\m^{-2/3}}$.
To emulate a realistic scenario, the receiving aperture is positioned at $z=$ \qty{100}{\km} and has a diameter $A_{RX} = $ \qty{10}{\cm}.   
In the transverse plane, for each realization, the aperture can be either on the propagation axis or moved by the beam wander $r_c$ to the location of maximum received power, as if tip-tilt is compensated.
\begin{table}[t]
\large
    \centering
\begin{tabular}{ll|rl|rl}
\hline \hline
                    &                        & \multicolumn{2}{c}{Receiver on axis}                                & \multicolumn{2}{c}{Beam wander corrected}                                                  \\
\textbf{Beam Class} & \textbf{$\mathbf{r_c}$ {[}m{]}} & \textbf{P$_{RX}$ {[}dBm{]}} & \multicolumn{1}{l}{\textbf{$\mathbf{\sigma_I^2}$}} & \multicolumn{1}{l}{\textbf{P$_{RX}$ {[}dBm{]}}} & \multicolumn{1}{l}{\textbf{$\mathbf{\sigma_I^2}$}} \\ 
\hline
OPB $\gamma = 0.1$  & 0.23                   & 7.00                       & 0.3448                                 & 9.67                                           & 0.1254                                 \\
OPB $\gamma = 1$    & 0.15                   & 13.45                      & 0.3700                                 & 17.87                                          & 0.2038                                 \\
OPB $\gamma = 1.9$  & 0.20                   & 13.28                      & 0.4337                                 & 18.17                                          & 0.2175                                 \\
Collimated GB       & 0.45                   & 4.67                       & 0.8274                                 & 12.32                                          & 0.1651                                 \\
Focused GB          & 0.23                   & 15.92                      & 1.0894                                 & 21.87                                              & 0.1859 
                                \\     
\hline \hline
\end{tabular}
    \caption{Turbulent propagation: average beam wander $r_c$, average power received $P_{RX}$ and scintillation index $\sigma_I^2$ by a \qty{10}{\cm} aperture, w/ and w/o beam wander correction at $z =$ \qty{100}{\km}.
    }
    \label{tab:statistics}
\end{table}
A focusing ($\gamma=1.9$) OPB exhibits a gain of more than \qty{5.8}{\decibel} in link budget compared to a collimated GB with beam wander correction and \qty{8.6}{\decibel} without it. Compared to a focused GB, the received power for $A_{RX}=$ \qty{10}{\cm} is at least \qty{2.6}{\decibel} lower. These findings are in agreement with similar simulations\cite{hu_experimental_2022}. The advantage of OPBs over focused GB is their self-healing property: in absence of correction, the scintillation index of any OPB is more than halved compared to focused GBs.

\section{Conclusions}
In the context of a \qty{10}{\km} high, \qty{100}{\km} long air-to-air link, 
we confirmed the improvement in link budget by employing optical pin beams instead of Gaussian beams by numerical propagation under the assumption of constant turbulence strength, $C_n^2 = \qty[exponent-product=\cdot]{1.66e-17}{\m^{-2/3}}$.
Given the same transmitting aperture, receiving aperture and transmitted power, on axis OPBs' received power is up to \qty{8.6}{\decibel} greater than collimated GBs and the scintillation index is halved. 
Moreover, OPBs exhibit a focusing region spanning tens of kilometers, an advantage over focused GBs in case of a moving target.
However, we demonstrated the necessity of large transmitting apertures in order to achieve long propagation distances and that their relationship is quadratic. 
Specifically, for an OPB with $\gamma = 1$, the minimum diameter needed is \qty{14}{\cm} if the target is positioned at a distance of \qty{1}{\km} and \qty{1.4}{\m} if it is \qty{100}{\km}. 
Greater power law exponents (focusing OPBs) require even larger apertures. 
This is a demanding requirement, as apertures greater than \qty{10}{\cm} are impractical to implement on air- or space- borne platforms of any size. 
OPBs may be more suitable for up-link applications from ground stations; in this scenario new simulations with a variable structure coefficient are to be addressed. 
While for the investigated class of OPBs, it is shown that very large apertures are required for long link distances, 
future research may determine if this relationship holds in the same way for other kinds of beams that exhibit non-diffracting and self-healing properties. 

\acknowledgments 

The authors thank SPIE for their support in waiving the conference fees and providing financial support for travel and lodging costs in order to present this work.

\bibliography{report} 
\bibliographystyle{spiebib} 

\newpage
\appendix    
\section{Propagation through atmospheric turbulence}
\label{sec:app_A}
\begin{figure}[h!]
     \centering
     \begin{subfigure}[b]{0.95\textwidth}
         \centering
         \includegraphics[width=1\textwidth]{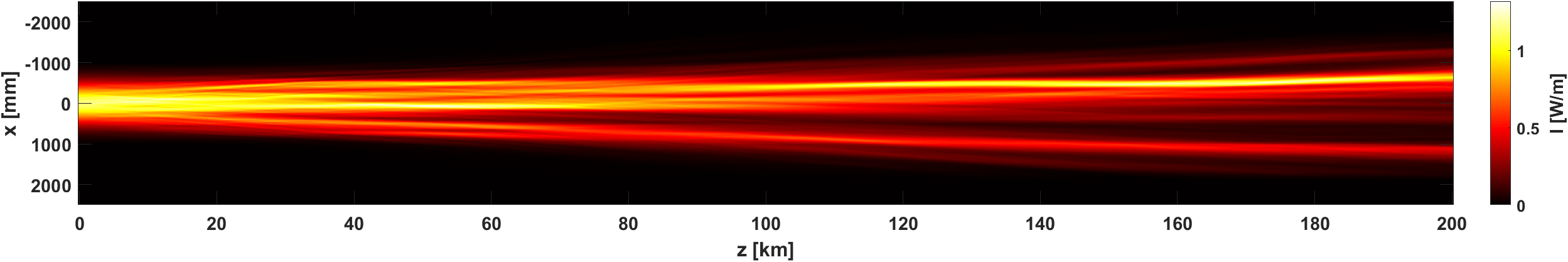}
         \caption{Collimated GB with $R \rightarrow \infty$}
         \medskip
     \end{subfigure}
     \hfill
     \begin{subfigure}[b]{0.95\textwidth}
         \centering
         \includegraphics[width=1\textwidth]{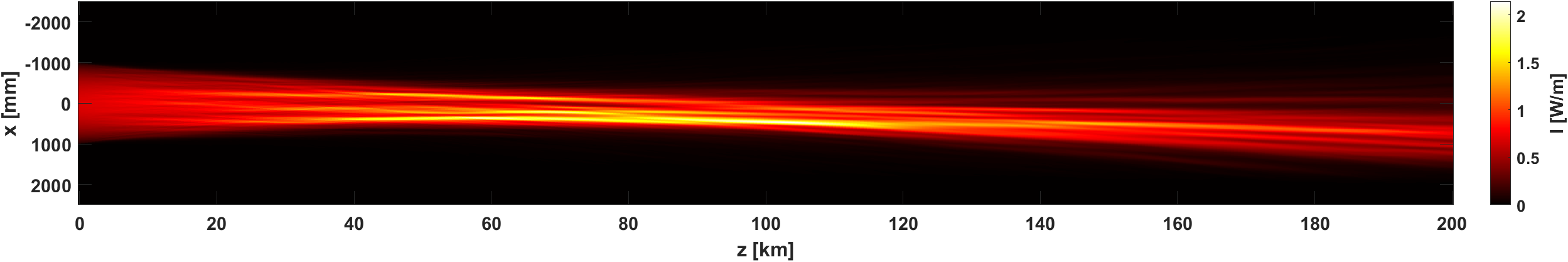}
         \caption{Focused GB with $R=\qty{100}{\km}$}
         \medskip
     \end{subfigure}
     \hfill
     \begin{subfigure}[b]{0.95\textwidth}
         \centering
         \includegraphics[width=1\textwidth]{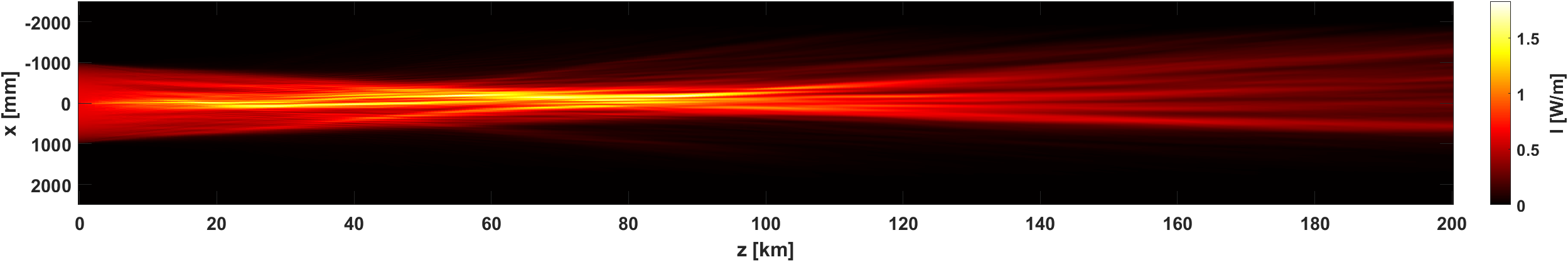}
         \caption{OPB with $\gamma=$ \qty{0.1}{}, $z_m =$ \qty{125}{\km}}
         \medskip
     \end{subfigure}
          \hfill
     \begin{subfigure}[b]{0.95\textwidth}
         \centering
         \includegraphics[width=1\textwidth]{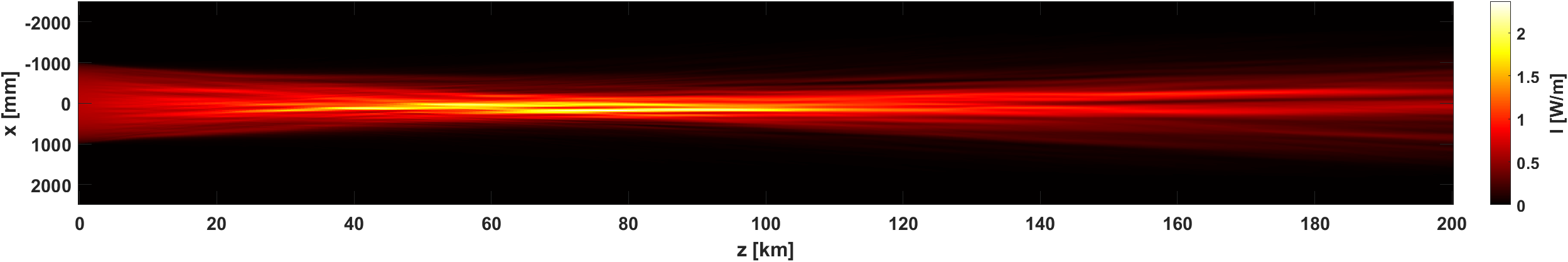}
         \caption{OPB with $\gamma=$ \qty{1}{}, $z_m =$ \qty{125}{\km}}
         \medskip
     \end{subfigure}
          \hfill
     \begin{subfigure}[b]{0.95\textwidth}
         \centering
         \includegraphics[width=1\textwidth]{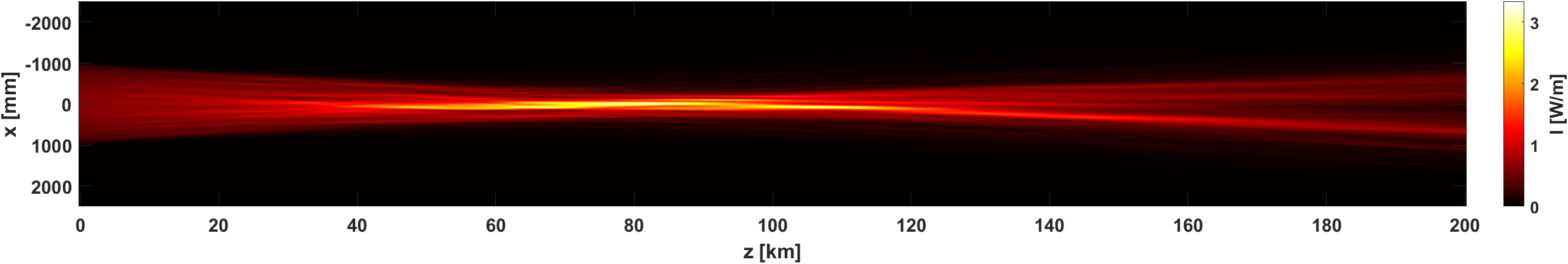}
         \caption{OPB with $\gamma=$ \qty{1.9}{}, $z_m =$ \qty{125}{\km}}
         \medskip
     \end{subfigure}
        \caption{Intensity profile of different beam types in atmospheric turbulence with $C_n ^2 = $ \qty[exponent-product=\cdot]{1.66e-17}{\m^{-2/3}}. All beams have the same transmitting aperture $A_{TX}=$ \qty{2}{\m} and transmitted power $P_{TX}=$ \qty{1}{\watt}.}
        \label{fig:intesity_profile_with_turbulence}
\end{figure}

The intensity profiles of OPBs with different power law exponents ($\gamma$) propagating in vacuum, over meter-scale distances, are available in related works \cite{li_direct_2020}.
Figure \ref{fig:intensity_profile} showed the propagation of an OPB (with $\gamma = $ \qty{0.1}{}) from our simulations. Propagation of collimated and focused GBs under the same condition has not been reported, as they do not contribute to new insights. 
Instead, Figure \ref{fig:intesity_profile_with_turbulence} presents the first simulation results for OPB propagation under turbulent conditions, compared to GBs.

\begin{figure}[h!]
     \centering
     \begin{subfigure}[b]{0.48\textwidth}
         \centering
         \includegraphics[width=0.6875\textwidth]{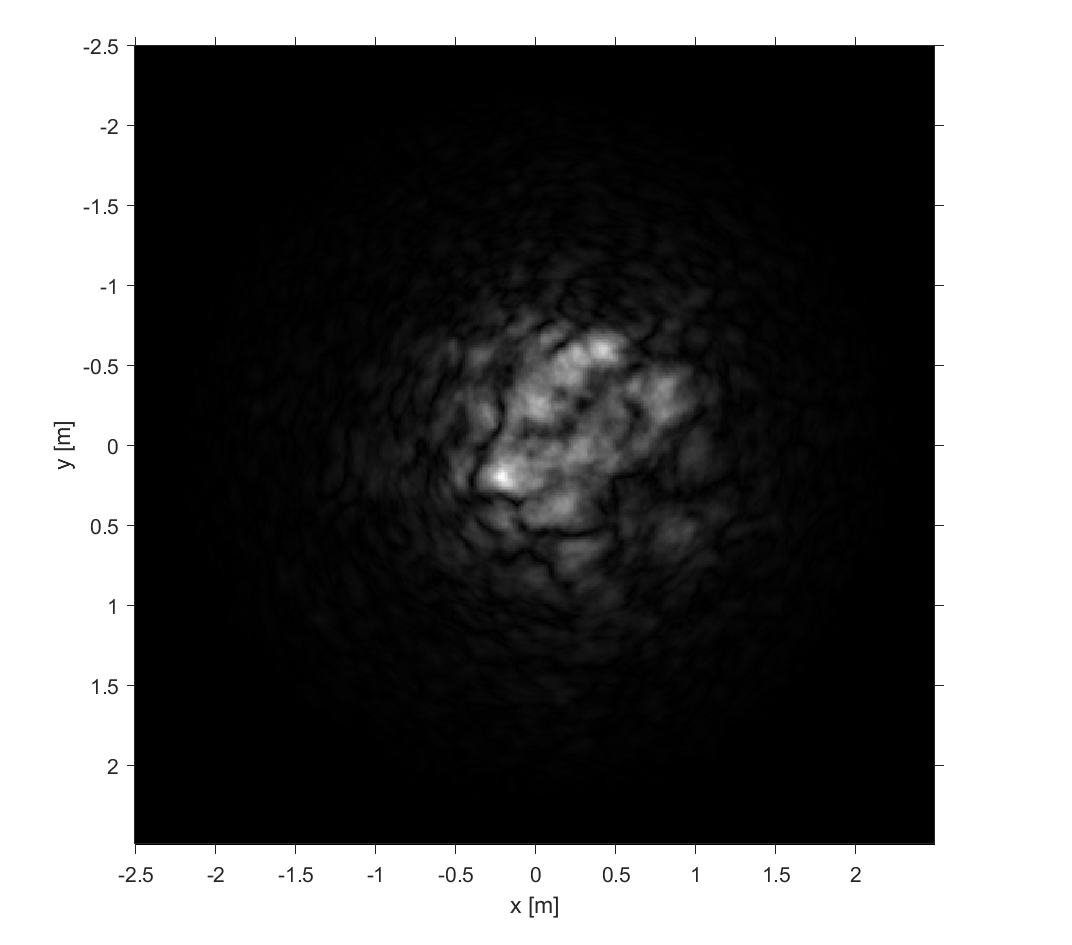}
         \caption{Collimated GB  with $R \rightarrow \infty$}
     \end{subfigure}
     \begin{subfigure}[b]{0.48\textwidth}
         \centering
         \includegraphics[width=0.6875\textwidth]{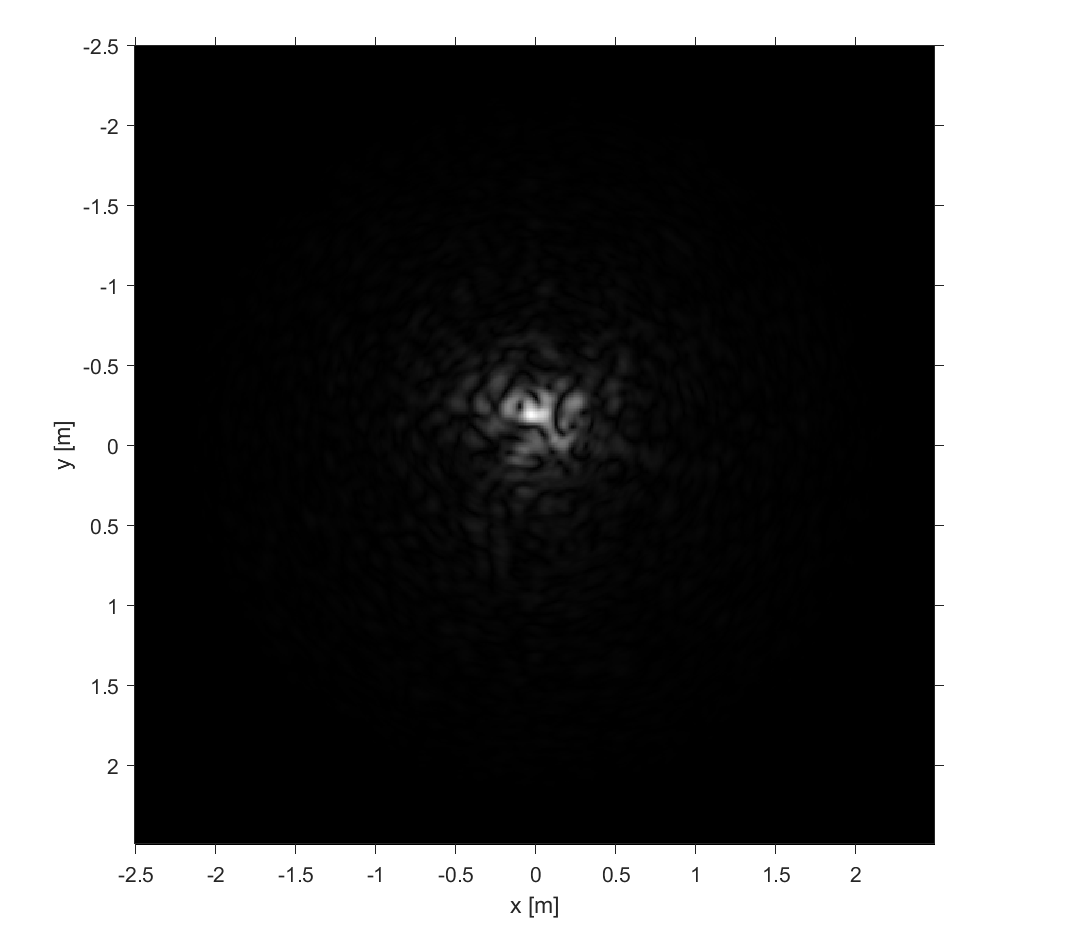}
         \caption{Focused GB with $R =$ \qty{100}{\km}}
     \end{subfigure}
     
     \medskip
     
     \hfill
     \begin{subfigure}[b]{0.32\textwidth}
         \centering
         \includegraphics[width=\textwidth]{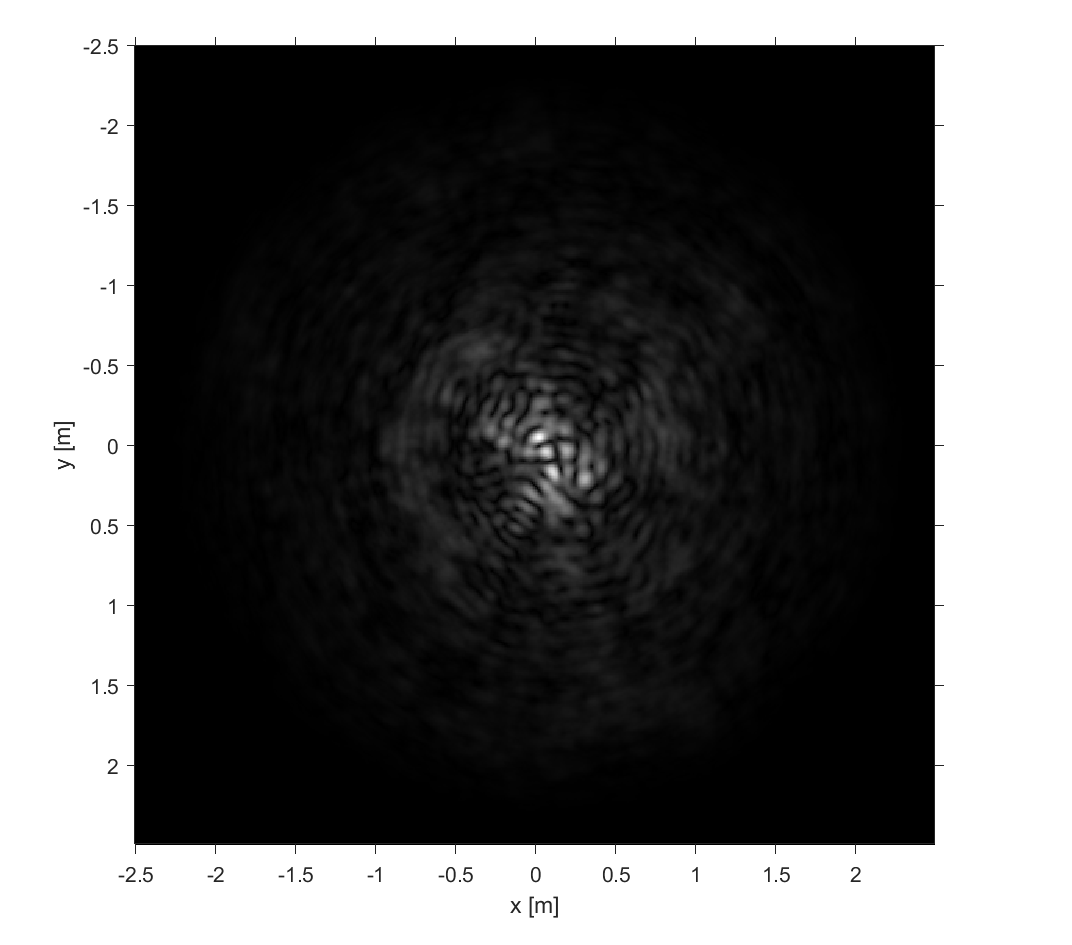}
         \caption{OPB with  $\gamma=$ \qty{0.1}{} }
     \end{subfigure}
          \hfill
     \begin{subfigure}[b]{0.32\textwidth}
         \centering
         \includegraphics[width=\textwidth]{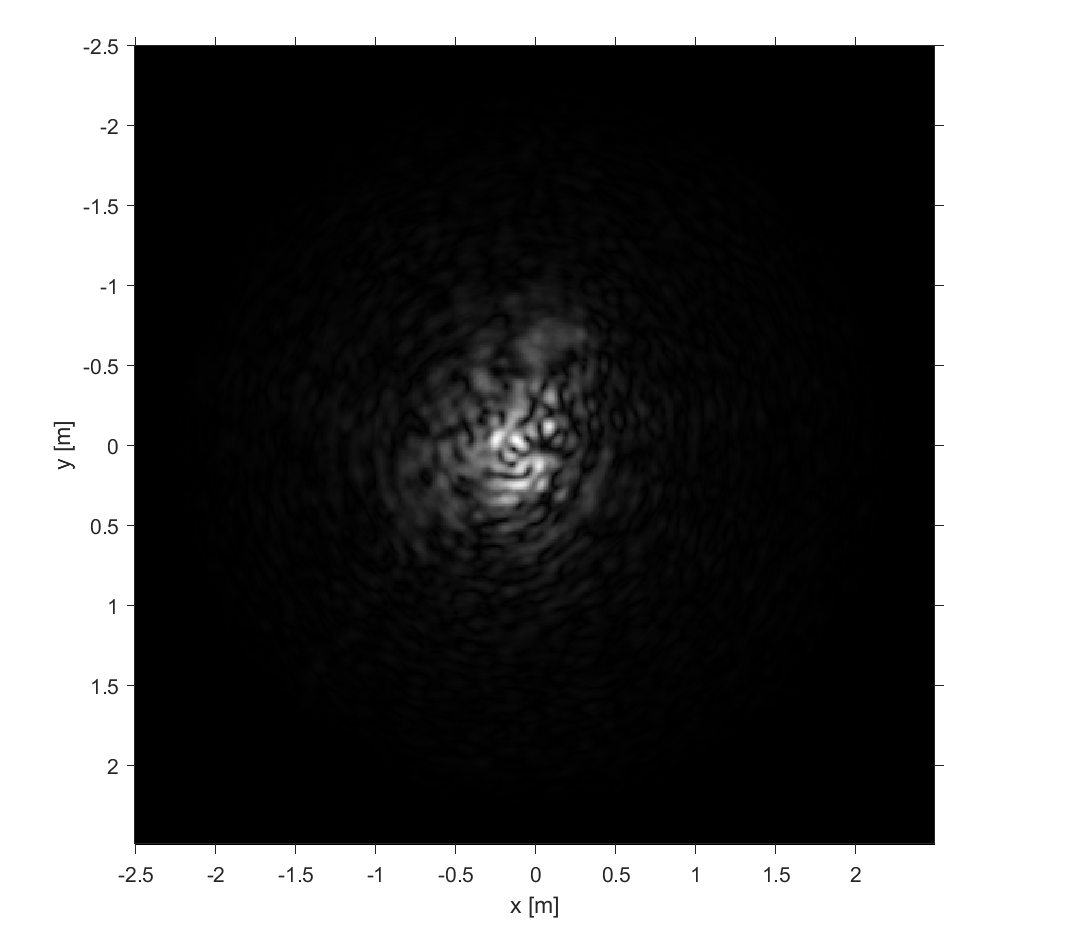}
         \caption{OPB with  $\gamma=$ \qty{1}{} }
     \end{subfigure}
          \hfill
     \begin{subfigure}[b]{0.32\textwidth}
         \centering
         \includegraphics[width=\textwidth]{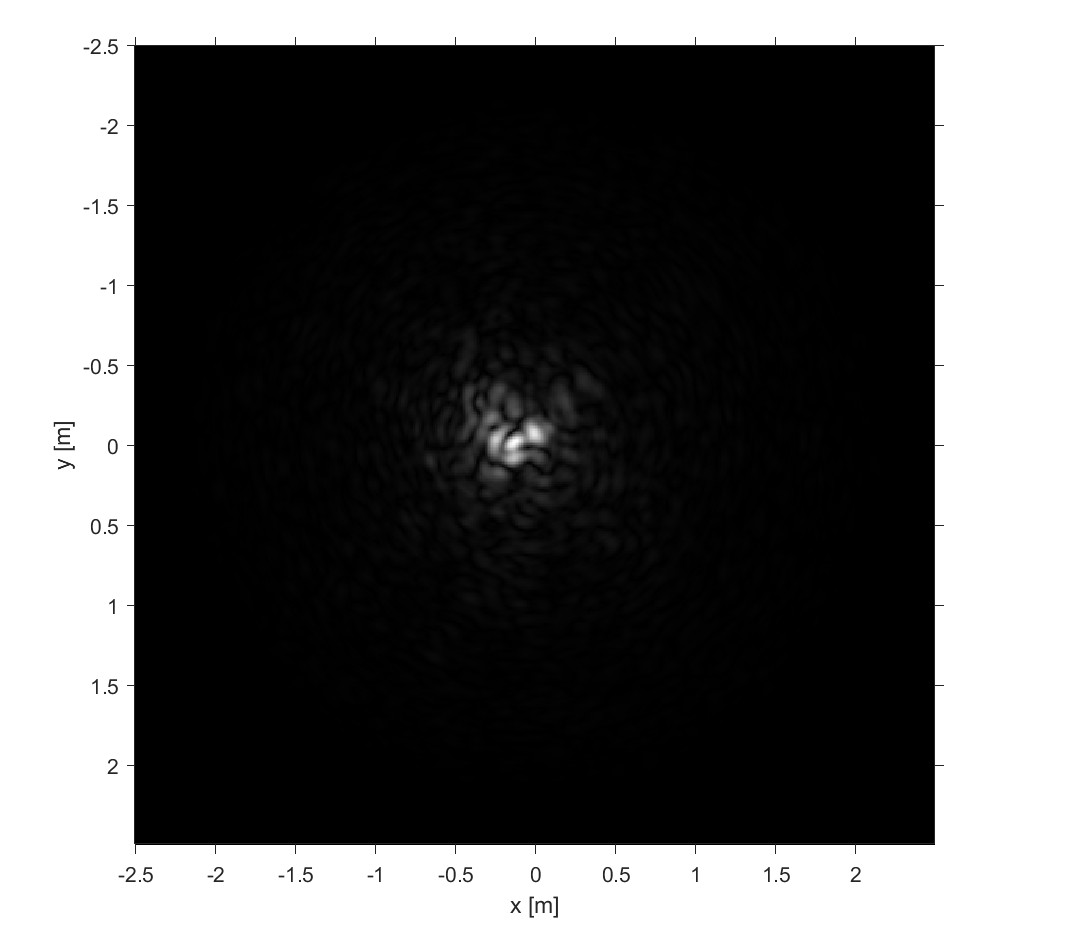}
         \caption{OPB with  $\gamma=$ \qty{1.9}{} }
     \end{subfigure}
     \medskip
        \caption{Comparison of the field amplitude of different beams at a distance $z=$ \qty{100}{\km} under constant turbulence $C_n ^2 = $ \qty[exponent-product=\cdot]{1.66e-17}{\m^{-2/3}}.  $A_{TX}=$ \qty{2}{\m}, $P_{TX}=$ \qty{1}{\watt}. }
        \label{fig:field_amplitude_100km}
\end{figure}

From our simulations, it may also be insightful to visualize the intensity "head-on", i.e. transversely, relative to the propagation axis (the $z$-axis). In this case, we have chosen to display the transverse plane (spanning the $x$- and $y$-axes) at the target distance where the receiving aperture is located. Figure \ref{fig:field_amplitude_100km} shows the field intensity (the square of the complex field amplitude) calculated on a $1024 \times 1024$ pixel quadratic grid with a \qty{5}{\m} long side, at a distance of \qty{100}{km} under atmospheric turbulence conditions.

\end{document}